\documentclass[%
 aip,
 amsmath,amssymb,
preprint,%
]{revtex4-1}

\usepackage{graphicx}% Include figure files
\usepackage{dcolumn}% Align table columns on decimal point
\usepackage{bm}% bold math

\usepackage{graphicx}
\usepackage{subcaption}
\usepackage{mwe}
\usepackage{caption}

\usepackage[utf8]{inputenc}
\usepackage[T1]{fontenc}
\usepackage{mathptmx}
\usepackage{xcolor}

\begin{document}

%\preprint{AIP/123-QED}

\title[Betatron radiation emitted during the direct laser acceleration of electrons in underdense plasmas]{Betatron radiation emitted during the direct laser acceleration of electrons in underdense plasmas}
% Force line breaks with \\

	\author{R. Babjak}
	\email[]{robert.babjak@tecnico.ulisboa.pt}
	%\thanks{}
	\affiliation{GoLP/Instituto de Plasmas e Fusão Nuclear, Instituto Superior Técnico, Universidade de Lisboa, Lisbon, 1049-001, Portugal}
	\affiliation{Institute of Plasma Physics, Czech Academy of Sciences, U Slovanky 2525/1a, 182 00 Praha 8, Czechia}

	\author{M. Vranic}
	\affiliation{GoLP/Instituto de Plasmas e Fusão Nuclear, Instituto Superior Técnico, Universidade de Lisboa, Lisbon, 1049-001, Portugal}

\date{\today}% It is always \today, today,

\begin{abstract}

Relativistic laser pulses can accelerate electrons up to energies of several GeV during the interaction with gaseous targets through the direct laser acceleration (DLA) mechanism. While the electrons are accelerated to high energies, they oscillate transversely to the laser propagation direction, emitting radiation. We demonstrate using particle-in-cell (PIC) simulations that the high accelerated electron charge enables DLA sources to emit  $\sim 10^{10}~\rm{photons}/0.1\%\rm{BW}$ at energies of hundreds MeV when interacting with multi-petawatt laser pulses. We provide an analytical estimate of the expected critical frequency for the DLA betatron spectrum which is in strong agreement with PIC simulations. We also show that using gas jets of low density ($\sim 10^{19}~\rm{cm^{-3}}$) is beneficial for the brightness of the source, since low plasma density produces collimated radiation. If the laser pulse is focused to an optimal spot size that results in the highest cut-off energies, conversion efficiency from laser to radiation can reach up to a few percent, which makes the DLA a promising high-brilliance source of gamma-ray radiation.

\end{abstract}

\maketitle

%\tableofcontents

\section{Introduction}

High-power lasers are capable of accelerating particles to high energies while interacting with various types of plasmas \cite{faure2004,geddes2004,mangles2004, fuchs2006,ziegler2024}. If the accelerated particles are electrons, they emit in X-ray and gamma-ray part of the spectrum while oscillating in strong plasma fields\cite{rousse2004,corde2013,cole2015}. The primary benefit of laser-plasma-based sources is the possibility of a high particle acceleration gradient, leading to improved compactness compared to conventional accelerators. New laser facilities operating in the multi-petawatt regime\cite{zou2015,webber2017,tanaka2020,yoon21} will substantially increase particle energies, unlocking potential for numerous new applications. Even though the laser-based sources allow for acceleration at shorter distances, there is a lot of room for improvement in terms of beam quality, conversion efficiency, and photon energies. 

The most extensively studied electron acceleration mechanism is laser wakefield acceleration (LWFA) which can provide mono-energetic electron beams with energies of several GeV\cite{leemans2014,gonsalves2019,aniculaesei2023}. While these electrons are being accelerated in the field of a plasma wave generated behind the laser pulse, they can emit radiation in the X-ray range\cite{wang2002, kostyukov2003,albert2014}. Emission of additional photons with even higher energies can be achieved when an energetic electron beam collides with the counter-propagating relativistic laser pulse and emits via non-linear Compton scattering \cite{harteman2007, thomas2012, phuoc2012,mirzaie2024,meir2024}. Another possibility is to collide an accelerated electron beam with a high-Z target which results in the emission of bremsstrahlung radiation\cite{perry1999, norreys1999, gunther2022, martinez2022}. Laser-produced radiation has a wide range of medical, biological, defense and industrial applications, but they are also important for condensed matter and high energy density science\cite{albert2016}.

An alternative approach to produce relativistic electron bunches and consequent betatron radiation is the direct laser acceleration mechanism\cite{pukhov1998,pukhov1999}. This is possible if the laser pulse is sufficiently long (hundreds of fs) to enable oscillations of electrons simultaneously in the field of an ion channel and a laser field. Energy transfer to electrons happens directly from the laser via resonant oscillations. The charge of accelerated electron bunch is generally in hundreds of nC per shot, with a wide energy distribution and cut-off energies up to several GeV\cite{babjak2024}. This has potential for applications where a high total charge of accelerated electrons is advantageous since the DLA can provide about a thousand times more charge compared to the LWFA\cite{hussein2021,shaw2021}. For example, using a DLA bunch can be used to generate a high yield of photoactivated neutrons\cite{pomerantz2014, gunther2022, cohen2024}. Also, during the interaction of relativistic electron bunch with a high-Z target, electron-positron pairs can be generated via Bethe-Heitler process for purposes of generating quasi-neutral electron-positron beams or to seed future coliders with positrons \cite{arrowsmith2024,streeter2024,warwick2017,sarri2015,sarri2022}. Similar to the LWFA scenario, when DLA accelerated electrons oscillate in the ion channel field, they become a high-yield source of betatron radiation\cite{chen2013,jansen2018,rosmej2021,gyrdymov2024,huang2016,shou2023,nakamura2012,lezhnin2018,stark2016,wang2020,kneip2008,rinderknecht2021, tangtartharakul2025,yeh2024}. High-charge of DLA electrons can also increase the efficiency of ion acceleration through the interaction with near-critical foams or exponential preplasma\cite{nuter2008,raffestin2021,bin2018}, where it is a possible mechanism of absorption together with a stochastic heating\cite{mendoca1983, paradkar2011,babjak2021}. High conversion efficiency into dense electron bunch can also be beneficial for seeding electron-positron pair cascades\cite{bellkirk2008,nerush2011,grismayer2017} in the regime when ultra-relativistic laser pulses collide and create a standing wave. A high number of energetic gamma-rays might also enable the experimental observation of the linear and non-linear Breit-Wheeler process\cite{chen2009,vranic2018,he2021,he2021_2,amaro2021,martinez2022,sugimoto2023}. 

Two main approaches to accelerate electrons by the DLA are possible: using underdense plasma that can be generated either by gas jets or using near-critical density targets that rely strongly on a relativistic transparency regime. The underdense regime can be also realized by expanding plasma in front of a thin foil and letting a laser pulse propagate through it. While near-critical density targets provide a higher charge of accelerated electrons and consequently higher radiation yield, they suffer from high divergence and low energies of electrons (few hundreds of MeV). Utilizing low plasma density enables reaching higher energies, which improves the beam divergence. Since low-density plasma allows longer propagation distances compared to near-critical density targets, continuous injection of electrons over several millimeters of propagation enables even undercritical density targets to produce high-charge bunches\cite{hussein2021, shaw2021, babjak2024}. In both regimes, electrons exceeding the vacuum energy limit in the DLA regime have been observed experimentally \cite{gahn1999,mangles2005,kneip2008,willingale2013,rosmej2019,rosmej2020,willingale2018,hussein2021,rinderknecht2021,lemos2024}. Recently, the optimal focusing strategy that results in the highest achievable electron energies \cite{babjak2024} has been observed in the experiment on 100 TW facility\cite{tang2024}.

\begin{figure}[h]   
    \centering
	\includegraphics[width=.79\textwidth]{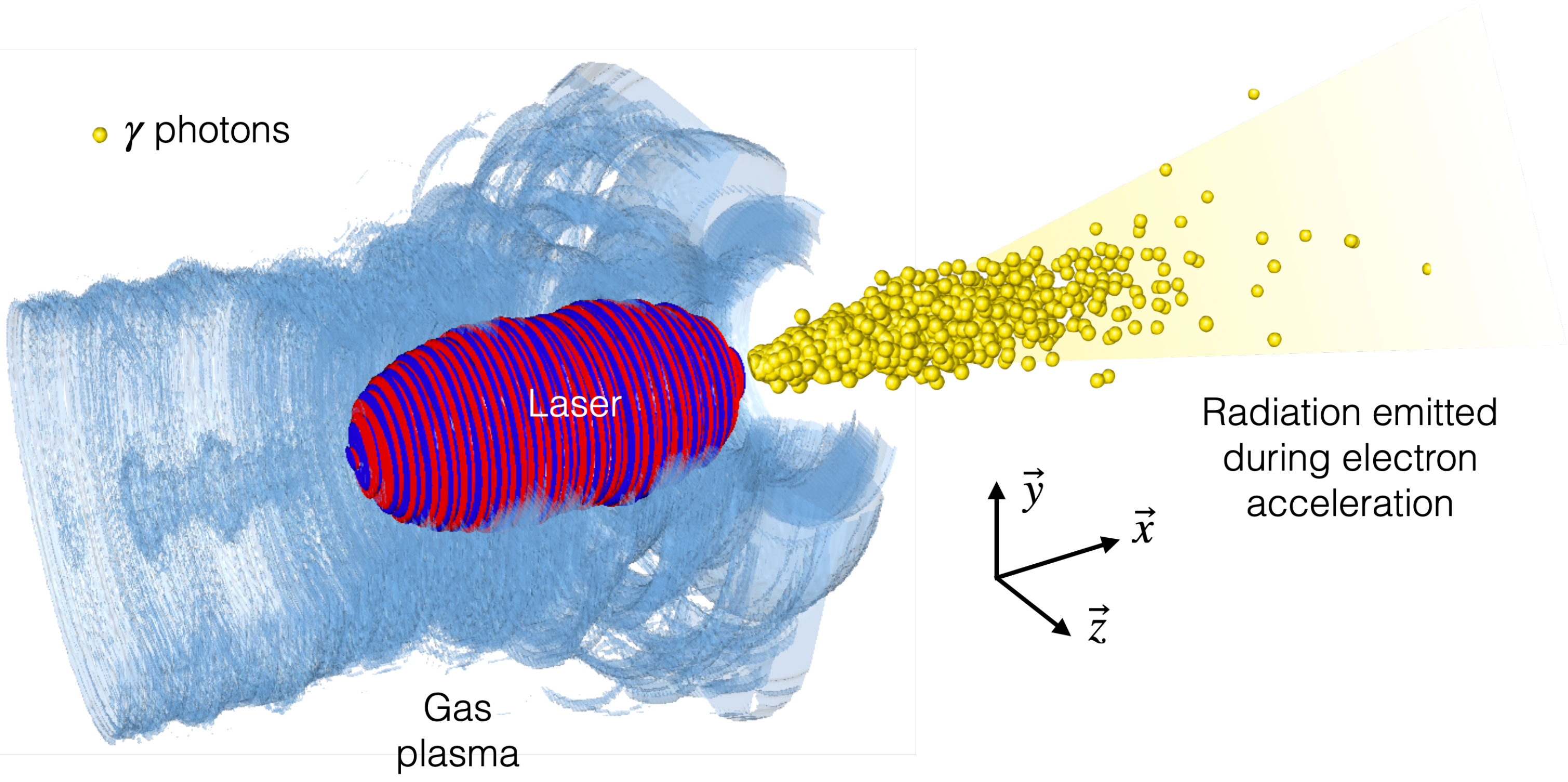}
	\caption{ Collimated photon emission during direct laser acceleration in a gas. An intense laser pulse propagating through the plasma target accelerates electrons to relativistic energies. Accelerated electrons co-propagate with the laser pulse and due to transverse betatron oscillations in the field of an ion channel, they emit collimated gamma-ray photons. }\label{fig:setup_3d}
\end{figure}

In this work, we demonstrate the potential of the DLA electrons accelerated in underctitical gas targets to serve as a collimated high-yield X-ray and gamma-ray source (see schematics in Fig. \ref{fig:setup_3d}). We show, that optimally accelerated DLA electrons can emit synchrotron radiation with critical energy exceeding hundred MeV using a 10 PW laser. The radiation is collimated within several tens of milirad with a total yield of up to $10^{10}$ photons per 0.1\% of the bandwidth (BW) near the critical energy. Such a source is tunable in terms of emitted radiation frequency (i.e. photon energy) since the synchrotron radiation spectrum depends on the background ion channel field strength, which can be controlled by changing the gas jet density. We provide an analytical scaling for the critical frequency of emitted radiation and we find an excellent agreement with Quasi-3D partice-in-cell (PIC) simulations. Our PIC simulations are performed using laser and plasma parameters in the optimal regime, which leads to the highest achievable electron energies for any given laser power\cite{babjak2024}, where we also observe agreement of achieved electron cut-off energies with the previous analytical predictions. The high efficiency of the DLA regime at densities between $10^{19}-10^{20} \rm{cm^{-3}}$ combined with the possibility to accelerate hundreds of nC of electrons to energies of several GeV opens an outstanding opportunity to construct a DLA-based high-yield collimated gamma-ray source. Furthermore, we observe in our PIC simulations that up to a few percent of laser energy can be converted into gamma radiation, which makes the DLA-based radiation source efficient compared to alternatives \cite{albert2016}.

\section{Analytical description}

We start the analytical description by summarizing the fundamentals of DLA acceleration in underdense plasmas in Section \ref{sec:dla_theor}, while the basic description of synchrotron emission is summarized in Section \ref{sec:syn_emiss}. Then estimate the properties of the radiation emitted during the acceleration process in Section \ref{sec:dla_rad} and present it as a function of what we know about the electron acceleration properties.

\subsection{DLA acceleration}\label{sec:dla_theor}
The DLA is a consequence of resonant coupled oscillations of electrons in the field of a laser pulse with a frequency $\omega_0$ and betatron oscillations in the field of an ion channel with frequency $\omega_{\beta}=\omega_p/\sqrt{2\gamma}$\cite{pukhov1999,pukhov2002, khudik2016,arefiev2016,arefiev2016_2,vranic2018b}. Ion channel fields consist of radial electric field component $E_r$ present due to the charge separation and azimuthal magnetic field $B_{\varphi}$ generated by the electron current. Both fields act on electrons providing a force pushing them towards the channel axis. The ratio between the electric and magnetic component of focusing ion channel forces depends on the plasma density but does not significantly influence the electron dynamics since the sum of both forces acting on relativistic electron co-propagating with the laser can be expressed as

\begin{equation}\label{eq:chan_force}
    F_C=\frac{m_e\omega_p^2}{2}r.
\end{equation}

Electrons that satisfy the resonant condition \cite{khudik2016,arefiev2014, arefiev2012} start to gain energy through DLA. When the background plasma density is constant, electrons oscillate with a constant amplitude given by the initial conditions of electrons, which is a consequence of the canonical momentum conservation. The conserved quantity, the integral of motion for this setup can be derived \cite{arefiev2012}

\begin{equation}
    I = \gamma - \frac{p_x}{m_ec} + \frac{\omega_p^2y^2}{4c^2},
\end{equation}

where $\gamma$ is electron's Lorentz factor, $p_x$ is momentum in the laser propagation direction, $\omega_p$ is a plasma frequency and $y$ is the distance from ion channel axis. 

The resonant condition restricts the oscillation amplitude of electrons that can achieve the resonant acceleration. Electrons with oscillation amplitude $Y$ higher than $y_{\rm{max}}$ can not be resonantly accelerated by the most efficient first harmonic resonance. The value of the maximum resonant oscillation amplitude can be expressed as \cite{babjak2024}

\begin{equation}\label{eq:ymax}
    y_{\rm{max}}=\frac{2c}{\omega_p}\sqrt{\left(\frac{a_0\omega_p}{\omega_0 \varepsilon_{cr}} \right)^{2/3}-1},
\end{equation}

where $a_0$ is the dimensionless laser field amplitude and $\varepsilon_{cr}$ is the resonance parameter with value 0.2. 

The maximum energy that electrons can achieve for a given plasma density can be expressed as $\gamma_{\rm{max}} \approx 2 I^2\omega_0^2/\omega_p^2$, where $I$ is proportional to the oscillation amplitude $Y$\cite{khudik2016,jirka2020}. This implies that electrons further from the channel axis can reach higher energies. As a consequence of that, eletrons with $Y=y_{\rm{max}}$ have potential to achieve the highest energies. If the laser pulse is sufficiently wide ($W_0>y_{\rm{max}}$) and stable laser propagation is ensured for sufficient distance $L_{\rm{acc}}/\lambda=0.78a_0^{2/3}\varepsilon_{cr}^{-5/3}(n_e/n_c)^{-2/3}$, the maximum energy that electrons can reach is\cite{babjak2024}

\begin{equation}\label{eq:gmax_res}
    \gamma_{\rm{max}}^{res}=2 \left(\frac{a_0}{\varepsilon_{cr}}\right)^{4/3}\left(\frac{n_e}{n_c}\right)^{-1/3}.
\end{equation}

If the density of a background plasma changes during the laser propagation, the maximum energy of each electron is defined at the moment when it starts to be resonantly accelerated and depends on the conditions at the moment when resonant acceleration started. Therefore, the limit on the maximum energy that electrons can reach can still be calculated for gradient plasmas \cite{babjak2024b}.

\subsection{Synchrotron emission}\label{sec:syn_emiss}

When we know the properties of electrons while being accelerated by the DLA as well as how the electrons oscillating in an ion channel emit radiation, we can use it to describe the emission during the acceleration stage of DLA. Electrons accelerated by the direct laser acceleration gain energy in the fields of an ion channel and a laser, they undergo betatron oscillations with a frequency $\omega_{\beta}$. During oscillations, electrons emit betatron radiation in the electron propagation direction\cite{schwinger1949, wang2002, kostyukov2003, rousse2004, albert2014}. The instantaneous spectral emission rate of a classical synchrotron $dW/d\hbar \omega$ in case of high $\gamma$ can be expressed as\cite{jackson1999}
\begin{equation}
    \frac{dW}{d(\hbar \omega)} = \frac{\alpha}{\sqrt{3}\pi\hbar\gamma^2} \int_{\xi}^{\infty} K_{5/3}(y)dy, 
\end{equation}

where $\xi $ is the ratio of $\omega$ to critical frequency $\omega_c$ expressed as

\begin{equation}
    \xi = \frac{2\hbar \omega}{3\chi_e \gamma mc^2} = \frac{\omega}{\omega_c},
\end{equation}

where $K_\nu$ is a modified Bessel function of the order $\nu$. The $\alpha\simeq 1/137$ is the fine structure constant, $\hbar$ is the reduced Planck's constant and $\chi_e=\gamma B/B_{s}$ is the quantum non-linearity parameter that compares the field strength observed in the electron instantaneous rest frame with $E_{s}=B_{s}c = m^2c^3/e\hbar$ which is the critical field of quantum electrodynamics. The critical frequency is then $\omega_c=3\gamma^3\omega_0/2$, where the $\omega_0=eB/\gamma m$ is the cyclotron frequency. The radiated spectrum is broad with a peak located at frequency $0.29\omega_c$. At frequencies higher than the critical frequency, the number of photons rapidly drops and the shape of the spectrum can be approximated as $\sim \sqrt{\omega/\omega_c}\exp(-\omega/\omega_c)$. This can be extended to a weakly quantum regime using Fermi–Weizäcker–Williams method\cite{lieu1993,martins2016}. When we consider an electron propagating through the ion channel with focusing forces linearly increasing with the distance from the channel axis $r$ that depends on the background plasma density $n_e$, the critical frequency can be estimated as\cite{jackson1999, rousse2004}

\begin{equation}\label{eq:ec_SI}
    \hbar \omega_c [\rm{eV}] = \frac{5.8}{\lambda[\rm{\mu m}]}  \frac{n_e}{n_c}\frac{r}{\lambda}\gamma^2.
\end{equation}

\subsection{Radiation emitted by the electrons undergoing DLA}\label{sec:dla_rad}

When electrons are accelerated by the DLA, their cut-off energy increases with the propagation distance. The total radiated spectrum is a superposition of radiation emitted by an ensemble of particles with broad energy distribution characteristic for DLA, that evolves over interaction time. The oscillation amplitude of electrons is also not identical for all electrons but varies depending on initial conditions during the injection process, and consequently the value of the integral of motion $I$. This influences the value of the focusing field at the moment of emission. Such complexity of the accelerating process makes it very challenging to describe the full spectrum of radiation emitted during the DLA acceleration. However, it is possible to estimate the effective critical frequency of the radiation. To calculate the total emitted spectrum exactly, it would be necessary to account for the time evolution of the energy spectrum, and the distribution of all electrons according to the oscillation amplitudes of all electrons, which is currently unknown. To estimate the effective critical energy of the radiation we first assume the linear increase of the maximum energy in the distribution $\gamma(t)=t~\gamma_{\rm{max}}/T_{\rm{acc}}$, where $T_{\rm{acc}}$ is the acceleration time to reach energy $\gamma_{\rm{max}}$. To correctly describe at least the most energetic part of the spectrum, we assume that photons are emitted by the electrons with energies close to the cut-off electron energy $\gamma(t)$ at given time $t$ of the acceleration and that they radiate at the highest resonant distance from the axis given by Eq. (\ref{eq:ymax}). This approximation is reasonable because the tail of photon distribution is emitted by the most energetic electrons, which are the electrons with the maximum oscillation amplitude for DLA\cite{babjak2024}. For $a_0\omega_p/(\omega_0\varepsilon_{cr})\gg 1$ which is easily satisfied for high intensities, the maximum radial oscillation amplitude can be approximated as 

\begin{equation}\label{eq:ymax_approx}
   \frac{ y_{\rm{max}}}{\lambda} \approx \frac{1}{\pi}  \left( \frac{a_0}{\varepsilon_{cr}} \right)^{1/3} \left( \frac{\omega_p}{\omega_0}\right)^{-2/3}.
\end{equation}

Combining this with the Eq. (\ref{eq:chan_force}), it is possible to estimate the force acting on electrons at the moment of radiation. We also account for the average number of emitted photons at critical frequency, which scales with the electron energy as $\sim \sqrt{\gamma}$, since it is proportional to the wiggler strength parameter $K=2 \pi \gamma r_0 / \lambda_{\beta}\sim \sqrt{\gamma n_e}r_0$, where $r_0$ is the electron oscillation amplitude \cite{rousse2004}. We use the number of emitted photons as a relative weight in a weighted average of a critical frequency $\bar{E}_c$ estimate as $N_{ph}\approx \sqrt{\gamma/\gamma_{\rm{max}}}$. Then we get

\begin{equation}
    \bar{E}_c =\frac{1}{T_{\rm{acc}}} \int_0^{T_{\rm{acc}}} \frac{3eB}{2m}\gamma(t)^2 N_{ph}dt = \frac{3eB}{2T_{\rm{acc}}m\sqrt{\gamma_{\rm{max}}}}\int_0^{T_{\rm{acc}}} \left( \frac{t \gamma_{\rm{max}}}{T_{\rm{acc}}} \right)^{5/2}dt
\end{equation}

During the acceleration process, only the energy of the most energetic electrons $\gamma$ varies along with the number of emitted photons $n_{\rm{ph}}$, whilst $B$ stays constant due to the constant oscillation amplitude and the constant plasma background density. After integrating, we obtain

\begin{equation}
    \bar{E}_c = \frac{3eB}{2m}\gamma_{\rm{max}}^2 \times \frac{2}{7},
\end{equation}

which can be also understood as if we replaced the whole time-evolving spectrum by the radiation by a single electron with effective energy $\gamma_{\rm{eff}}=\sqrt{2/7}\gamma_{\rm{max}}=0.53\gamma_{\rm{max}}$. 

Substituting the $\gamma_{\rm{eff}}$ and $y_{\rm{max}}$ into Eq. (\ref{eq:ec_SI}), we obtain

\begin{equation}\label{eq:ec_total}
    E_c [eV] = \frac{2.1}{\lambda [\rm{\mu m}]} \left( \frac{a_0}{\varepsilon_{cr}}\right)^3
\end{equation}

as a critical frequency of the total emission radiated by an ensemble of electrons accelerated by the DLA during the acceleration to the highest achievable energy $\gamma_{\rm{max}}$. Note, that to satisfy assumptions of the derivation, sufficient propagation distance needs to be sustained and the laser spot size needs to be wider than $1.2y_{\rm{max}}$ \cite{babjak2024}. Also, the effects of radiation reaction are neglected. The fact that the critical frequency does not depend on the background density might be unexpected, but is consistent with the theory of DLA acceleration. It is true that critical frequency increases with plasma density for constant radial distance from the channel axis and electron energy, but in the DLA acceleration, lower plasma densities result in higher electron energies (see Eq. (\ref{eq:gmax_res})) and also allow for the acceleration and radiation further from the axis (see Eq. (\ref{eq:ymax_approx})), which enables the radiation in stronger background channel field. Both effects compensate for the decrease of plasma density in the case of optimal DLA. 

When electrons perform betatron oscillations, the radiation is emitted within a cone of an angle $K/\gamma$. Substituting energy achievable in the optimal DLA regime $\gamma_{\rm{max}}$ and the maximum oscillation amplitude $y_{\rm{max}}$, one obtains $K=2a_0/\varepsilon_{cr}$.  Thus, the emission angle $\Theta = K/\gamma$ can be expressed for the DLA emission as

\begin{equation}\label{eq:theta}
    \Theta = \left( \frac{n_e \varepsilon_{cr}}{n_c a_0} \right)^{1/3}.
\end{equation}

\section{Simulations of optimal direct laser acceleration of electrons }

The acceleration of electrons by the DLA can be enhanced significantly by applying an ideal focusing strategy \cite{babjak2024}. For each combination of laser power and background plasma density, the focal spot that enables reaching the highest possible cut-off energies can be found. Ideally, the spot size evolution should be constrained during the acceleration, which can be achieved using parabolic guiding channels that are now experimentally accessible\cite{oubrerie2022,levato2020,gupta2022}. The shape of a preformed guiding channel might influence the injection of electrons\cite{valenta2024}. Very efficient acceleration can be reached also in an unguided regime that relies on relativist self-focusing\cite{hussein2021}, but such a regime is less controllable and not fully explored yet.

\begin{figure}[h]   
	\includegraphics[width=.59\textwidth]{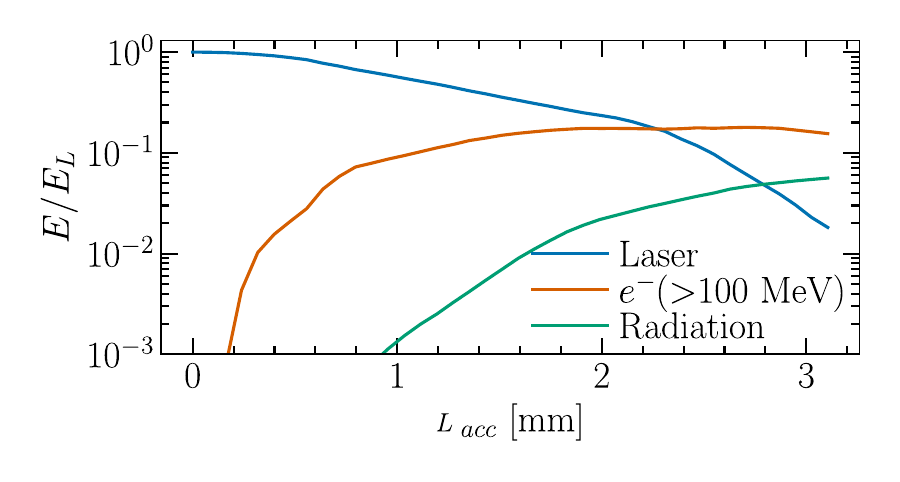}
	\caption{Energy transfer from the laser to electrons and radiation. The graph depicts the time evolution of the energy of the laser field, accelerated electrons and emitted photons during the simulation of 5 PW laser pulse propagating through the plasma channel with the density on the axis $0.05 ~n_c$. Only the energy of electrons with energies above 100 MeV is included. After 3 mm of propagation, the laser pulse is almost fully absorbed, 16 \% of laser energy was transformed into energetic electrons and 5 \% of laser energy was transferred into radiation. Energies are normalized by the initial energy of a laser pulse.   }
	\label{fig:time_evol}  
\end{figure}

We perform a set of Quasi-3D PIC OSIRIS\cite{fonseca2002,davidson2015} simulations for DLA acceleration in constant density plasma. The laser pulse spot size is chosen according to Eq. (4) in Babjak et al.\cite{babjak2024}, to ensure the optimal regime of electron acceleration. In our simulations, we initialize a parabolic quasi-neutral guiding plasma channel with a transverse density profile in a form $(n_w-n_e)\left(\frac{r}{R_{ch}}\right)^2 + n_e$, where $n_w$ is a channel wall density, $R_{ch}$ is a plasma channel radius and $n_e$ is a plasma density at the center of a channel. Laser pulse transverse profile is defined as $a_0(r) = a_0 \exp(-r^2/W_0^2)$, so the spot size $W_0$ corresponds to the distance where the field amplitude is $1/e$ of the maximum. The longitudinal laser envelope is chosen as polynomial function $a_0(\tau)=a_0\times [10\tau^3-15\tau^4+6\tau^5]$ expressing the first half of the envelope where $a_0$ increases, where $\tau = t/T_L$ and $T_L$ is a laser duration in FWHM of the laser field amplitude. The second part is symmetric around the point of the highest intensity, but instead with a decreasing field amplitude. This means that if we refer to the laser pulse with a duration $T_L$, the interval [$t_0-T_L/2$;$t_0+T_L/2$] around the laser center at $t_0$ contains 94\% of the laser energy. For such chosen laser pulse, the total energy can be expressed as $E[\rm{J}]=0.78\times P[\rm{PW}]\times T_L[\rm{fs}]$. The grid was discretized using 60 cells per wavelength in the laser propagation direction and 21 cells per wavelength in the transverse direction. The first two azimuthal modes were used for simulations. They account for the axisymmetric self-generated channel fields (mode m = 0) and for the non-axisymmetric linearly polarized laser field (m = 1). 32 particles per cell were used both for ion and electron species. We assume laser wavelength $\lambda= 1~\rm{\mu m}$ and the timestep $\Delta t = 25.4~\rm{as}$. 

\begin{figure}[h]   
    \centering
	\includegraphics[width=.59\textwidth]{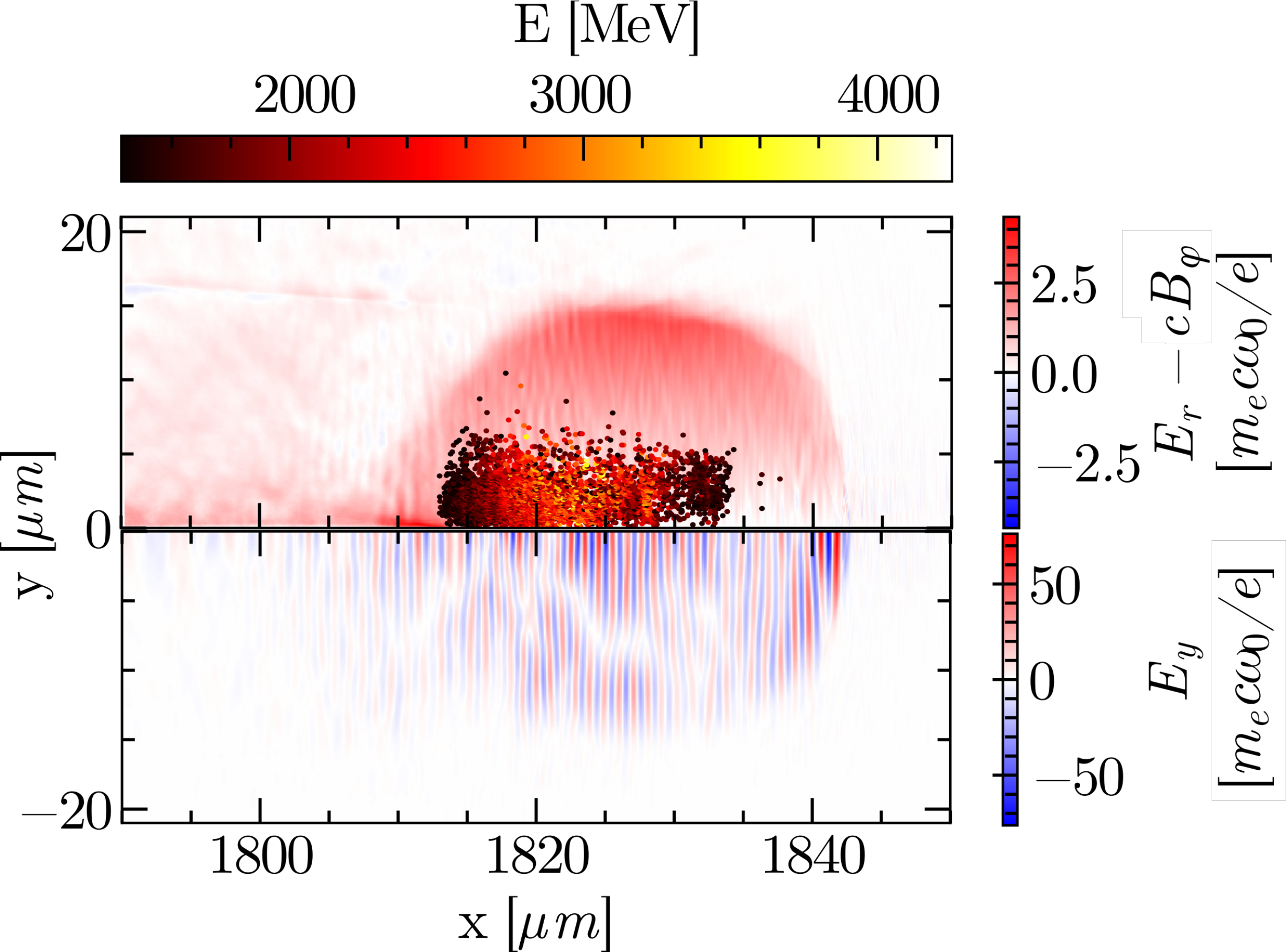}
	\caption{ Electrons accelerated by the DLA perform coupled oscillations in the field of an ion channel and a laser pulse. (a) Cylindrically symmetric ion channel focusing fields and the position of electrons during the acceleration. (b) Transverse component of oscillating laser electric field. Snapshot from the Quasi-3D PIC simulation of 5 PW laser pulse propagating through the plasma channel with the on-axis density $0.05 ~n_c$.  }\label{fig:setup}
\end{figure}

We run simulations until the time when accelerated electrons stop radiating. Laser and plasma parameters for selected simulations are given in Table \ref{tab:acc_params}. Laser power of 0.1, 1, 5 and 10 PW were investigated with various focusing geometries chosen. The propagation distance after which the highest energies were reached is referred to as $L_{acc}$, which is of the order of several mm. The Fig. \ref{fig:time_evol} shows the energy evolution during the simulation of a 5 PW laser propagating through the plasma channel with the on-axis density of $0.05~n_c$. It shows that the laser is almost fully depleted after 3 mm of propagation. After 2 mm, the total kinetic energy of electrons accelerated above 100 MeV does not significantly increase, but they continue radiating. At the end of the simulation 16 \% of laser energy is in energetic electrons, whilst 5 \% of laser energy was transformed into radiation. The radiation emission saturates once there is not enough of energy in the laser left to generate the ion channel fields that would enable electrons to radiate further. This hints that to ensure the efficient radiation, it is necessary to enable laser propagation not only long enough for electrons to gain energy, but also allow a prolonged oscillation period before the laser is fully depleted. The amount of radiated energy can be potentially increased if an electron bunch is sufficiently dense to drive its own wake, which might result in the radiation in an ion channel field also without the presence of the laser field \cite{clayton2009}. We observed conversion efficiencies from laser to electrons in tens of percent by a variety of simulations with different powers and densities. The same is true for conversion efficiency into radiation, which was in general of the order of a few percent.

\begin{table*}

\begin{ruledtabular}
\begin{tabular}{ccccccc}
  Laser power (PW) & Laser duration FWHM $a_0$ (fs) &  $W_0$ ($\mu m$) & $a_0$ & $R_{ch}$ ($\mu m$) &  $n_e$ ($n_c$) & $L_{acc}$ (mm)\\
 \hline
  0.15 & 350 & 5 & 16.7 & 25 & 0.03 & 2.0 \\
  1 (1) & 66.5 & 8 & 27 & 35 & 0.01 & 2.2 \\
  1 (2) & 250 & 4 & 54 & 25 & 0.1 & 1.2 \\
  5 & 150 & 8 & 60 & 25 & 0.05 & 1.8 \\
  10 (1) & 75 & 10 & 68 & 20 & 0.01 & 3.4 \\
  10 (2) & 150 & 7 & 97.14 & 14 & 0.05 & 2.4 \\

\end{tabular}

\end{ruledtabular}
\caption{ Laser and preformed plasma channel parameters used in our simulations chosen based on the optimal focusing strategy. $W_0$ is laser spot size at focus, $a_0$ is the peak laser normalized amplitude, $R_{ch}$ is preformed quasi-neutral parabolic channel radius, $n_e$ is the plasma density at the center of the channel and $L_{acc}$ is the propagation distance that was needed for electrons to reach the maximum achievable cut-off. }\label{tab:acc_params}
\end{table*}

\begin{figure}[h]
	\includegraphics[width=.59\textwidth]{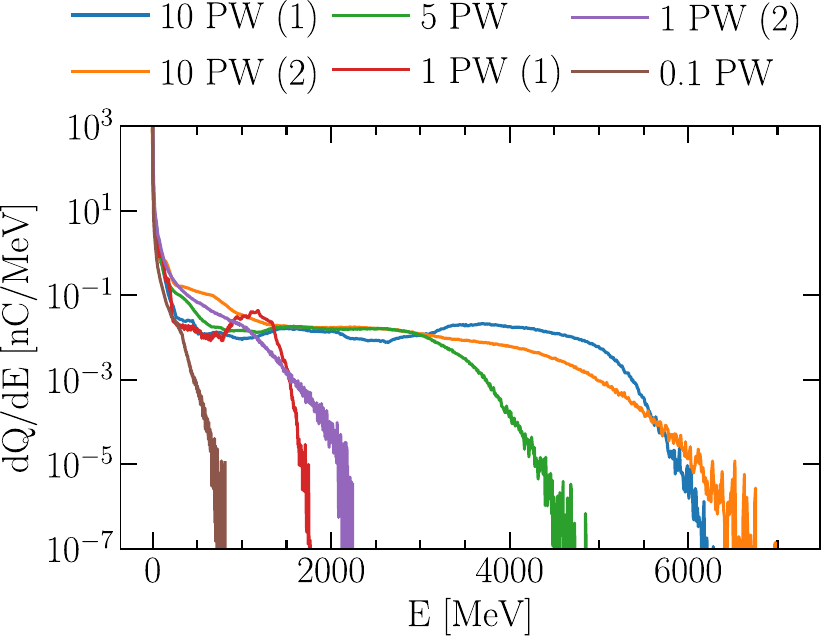}
	\caption{Laser energy can be converted into high-energy electrons with a broad energy spectrum with a conversion efficiency in tens of percent. Distributions are shown at the moment when they achieve the highest cut-off energies during the acceleration. For all simulations, the combination of laser spot size, power and plasma density was chosen according to the optimal focusing condition\cite{babjak2024}.}
	\label{fig:spectra}
\end{figure}

The dominant acceleration mechanism for electrons was the direct laser acceleration. This can be expected due to the presence of accelerated electrons in the field of a laser pulse and ion channel simultaneously, even though a bubble can be formed as in Fig. \ref{fig:setup}. Simultaneous action of both fields on electrons results in coupled oscillations that can accelerate them to over-ponderomotive energies after satisfying the resonant condition \cite{khudik2016,arefiev2012}. Another feature of the DLA is the broad energy spectrum shown in Fig. \ref{fig:spectra}. Even though the qualitative shape of spectra varies between different cases, the cut-off of the energy distributions agrees well with previously derived scaling for maximum energies achievable under optimal conditions \cite{babjak2024}. Approximately, for 0.1 PW laser the cut-off energy reached was 700 MeV, for 1 PW lasers 2 GeV and 5 PW laser delivered electron energies up to 4.5 GeV. The highest energy for 10 PW lasers exceeded 6 GeV, but the energy gain was limited either because of the radiation reaction limit imposed by the background channel fields \cite{jirka2020} or by the dephasing of a laser envelope and the electron bunch. Using longer laser pulse of 150 fs and low enough plasma density to suppress the effects of the radiation reaction (0.01 $n_c$) enables reaching 7.5 GeV energies. Nontrivial shapes of electron spectra ranging from Maxwellian-like to a combination of waterbag and Maxwellian tail distributions can be related to nonlinear interplay between plasma response to intense laser field along with laser pulse dynamics and needs further investigation. 

After the moment when the increase of electron energy cut-off saturates because the theoretical maximum was achieved\cite{babjak2024}, the energy conversion from laser to electrons can still be present. Continuing the acceleration after this moment might increase the total conversion efficiency, even though the energy cut-off does not increase. This is because the electrons that started the acceleration immediately at the beginning of the interaction had enough of time to reach the highest energies and they occupy the most energetic part of the spectrum. On the other hand, electrons that got injected at later times did not have enough time to reach the highest energies, and still occupy the low energy part of the spectrum. These electorns can be accelerated further. Continuous injection of electrons is one of the reasons why DLA provides a wide electron energy spectrum. 

\begin{table*}

\begin{ruledtabular}
\begin{tabular}{ccccc}
  Laser power (PW) & Laser duration & Ene. threshold $E_{th}$ (MeV) & Charge (nC) & Conv. eff (\%) \\
 \hline
  0.15 & 350 & 50 & 130 & 7 \\
  1 (1) & 66.5 & 100 & 66 & 44 \\
  1 (2) & 250 & 100 & 195 & 19\\
  5 & 150 & 200 & 330 & 17 \\
  10 (1) & 75 & 400 & 124 & 32 \\
  10 (2) & 150 & 400 & 198 & 15 \\

\end{tabular}
\end{ruledtabular}
\caption{\label{tab:ele_properties} Properties of accelerated electrons after the propagation of a laser for the distance $L_{acc}$ summarized in Tab. \ref{tab:acc_params}. Energy threshold is a minimum energy of electrons, that was accounted for the calculation of total accelerated charge and conversion efficiency.  }
\end{table*}

Properties of accelerated electrons with energies exceeding the energy threshold $E_{th}$ are shown in Table \ref{tab:ele_properties}. Data in the table correspond to electron spectra shown in Fig. \ref{fig:spectra}. Generally, optimal focusing strategy results in conversion efficiency of laser energy into energetic electrons in tens of percent, while total accelerated charge mostly exceeds 100 nC. Electron bunch divergence is observed to be in tens of milirad. 

\section{Simulated properties of emitted radiation}

\begin{figure}[h]
	\includegraphics[width=.59\textwidth]{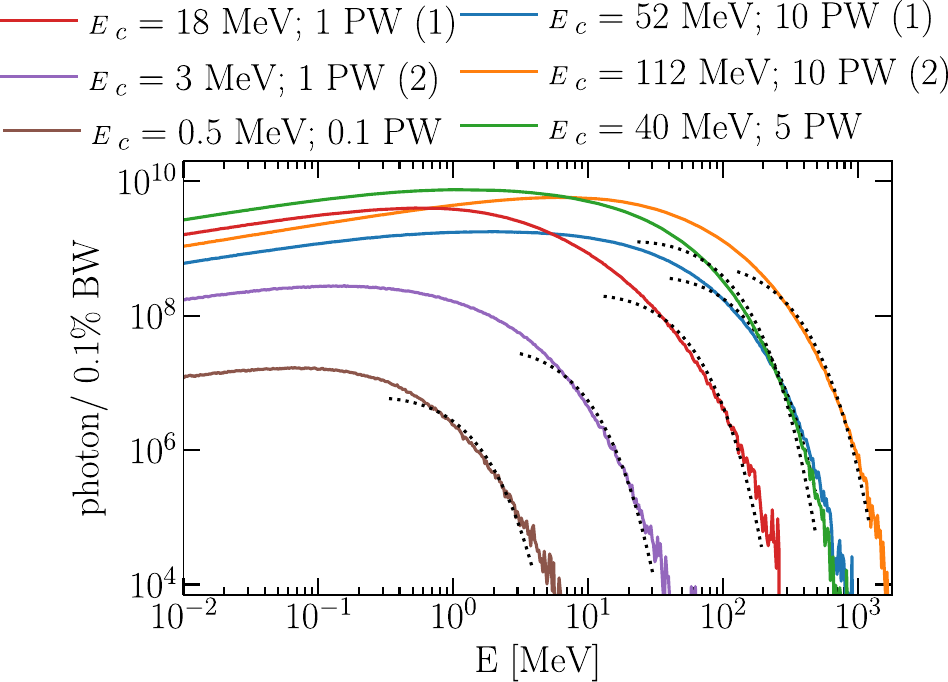}
	\caption{Cumulative emission radiated by accelerated electrons during the interaction of a laser with preformed parabolic plasma channel for lasers with various powers. Tails of the distribution can be described by the high-energy approximation of a synchrotron spectrum, $\sim \exp(-E/E_c)\sqrt{E/E_c}$, where $E_c$ stands for the critical frequency of a synchrotron radiation.}
	\label{fig:photon_spectra}
\end{figure}

As described above, radiation is emitted by the electrons during the acceleration stage and also partially after the maximum energy is reached during the oscillations in the ion channel fields. The radiated spectrum observed at the end of the simulations is shown in Fig. \ref{fig:photon_spectra}. Emitted radiation corresponds to the same simulations, for which the electron distribution functions were shown in Fig. \ref{fig:spectra}. Even though the total emitted radiation might have a complex shape due to reasons discussed earlier, it is possible to fit the tail of the spectrum using high-energy synchrotron approximation $A\sqrt{E/E_c}\exp(-E/E_c)$. This enables us to assign a critical frequency $\bar{E}_c$ for each simulation run. In general, we observe the increase in a number of emitted photons with laser power, since higher power lasers enable the acceleration of a higher number of electrons to higher energies and the number of photons emitted at critical frequency increases with the electron energy as $\sim \sqrt{\gamma}$. It is important to keep in mind that each simulation corresponds to the different combinations of $a_0$, $W_0$, and background plasma density $n_e$, see Table. \ref{tab:acc_params}. 

\begin{figure}[h]
	\includegraphics[width=.99\textwidth]{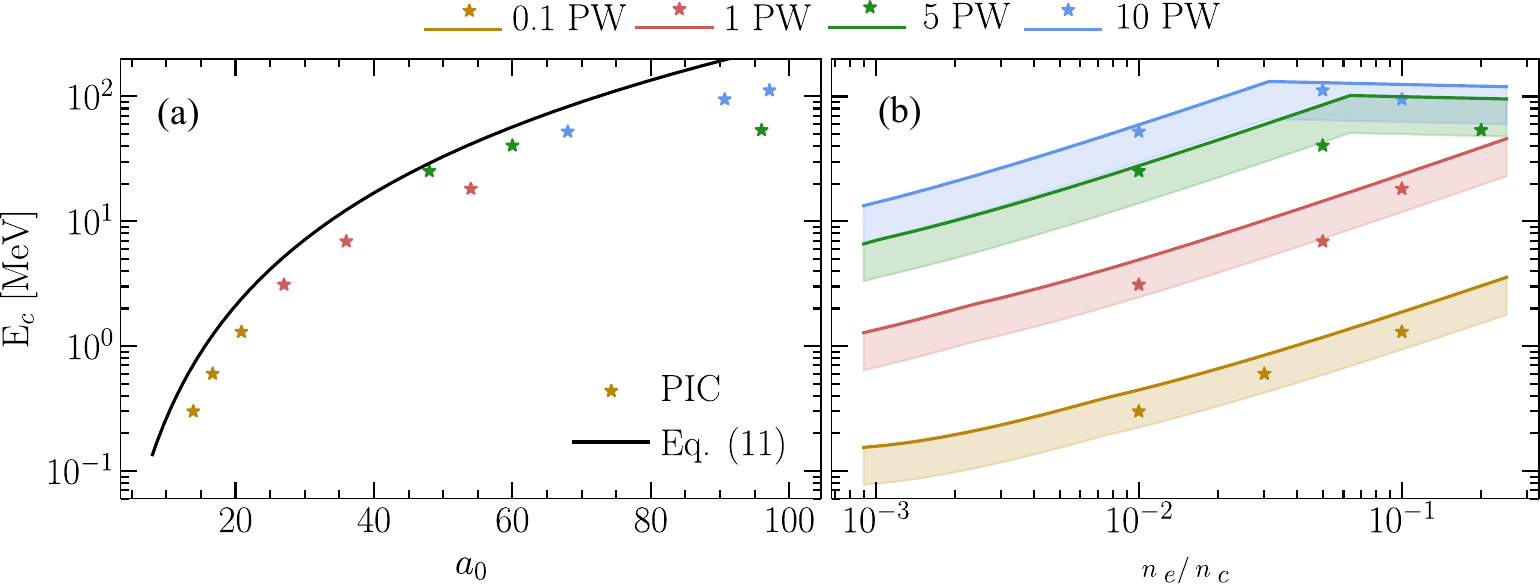}
	\caption{Scaling for the critical frequency of emitted radiation (Eq. \ref{eq:ec_total}) compared with values extracted from PIC simulations. (a) Critical frequency as a function of $a_0$. The background plasma density varies for each simulation according to the optimal focusing strategy. (b) Comparison of simulation results with the $E_c$ scaling for the highest achievable critical frequency for a given laser power and background plasma density, if optimal focusing is applied. The band represents the expected variation of $E_c$ if the maximum electron energy achieved is lower by 30 \% compared to the model. Eq. (\ref{eq:ec_total}) therefore gives excellent predictive capability for $E_c$. }
	\label{fig:ec_scaling_comparison}
\end{figure}

Critical frequencies extracted from PIC simulations can be compared with the analytical estimate for the critical frequency, Eq. (\ref{eq:ec_total}). Fig. \ref{fig:ec_scaling_comparison} (a) shows good agreement of simulations with the scaling for wide range of laser intensities. The comparison of simulation results and the scaling was possible because we made sure that the laser spot size is sufficiently wide and laser propagated long enough to allow electrons to reach the maximum predicted energy. In other words, the electron distribution function reached energies up to the predicted cut-off $\gamma_{\rm{max}}$ and electrons could radiate from the maximum resonant distance from the channel axis $y_{\rm{max}}$, which were the assumption for the derivation of Eq. (\ref{eq:ec_total}). Even though the critical frequency does not depend on the background plasma density, it is necessary to keep in mind that this is the case only because the ideal regime for the DLA in gas targets was reached\cite{babjak2024}. Without that, Eq. (\ref{eq:ec_total}) cannot be compared.

To better demonstrate that for each combination of laser power and background plasma density exists the maximum critical radiation frequency, we plot the same data as used in Fig. \ref{fig:ec_scaling_comparison} (a) as a function of background plasma density in Fig. \ref{fig:ec_scaling_comparison} (b). Data are compared with the theoretical estimate for the highest critical frequency for given laser power. To get such a scaling, we can estimate the value of $a_0$ resulting from the optimal focusing strategy (that was also used to choose our simulation parameters) using Eq. (4) in \cite{babjak2024}. The value of optimal $a_0$ for 0.1, 1, 5 and 10 PW was calculated, substituted into Eq. (\ref{eq:ec_total}) and compared with simulation results, see Fig. \ref{fig:ec_scaling_comparison} (b). The colored band under the line extends up to $0.5 ~E_c$ to account for the expected fluctuations in maximum electron energy up to $0.7\gamma_{\rm{max}}$ instead of $\gamma_{\rm{max}}$ due to deviations from ideal conditions. This can be expected in experiments, where initial parameters are not so easily controlled, but also in simulations, due to contributions of other processes that we neglected in the model. The line representing the scaling has a flat limit on the right-hand side of Fig. \ref{fig:ec_scaling_comparison} (b) for the case of 5 and 10 PW, due to the limit imposed by the radiation reaction (RR), originating from the background channel fields \cite{jirka2020,babjak2024}. This prevents electrons from reaching the highest achievable energy. Instead of $\gamma_{\rm{max}}$, we used $\gamma_{rr} \sim (\lambda a_0 \omega_0^2 /( \omega_p^2\sqrt{I}))^{2/5} $ for that section. Limited electron energy gain in the regime of higher plasma densities results in the limited increase of synchrotron critical frequency. Please note that PIC simulation data in Fig. \ref{fig:ec_scaling_comparison} (a) and (b) are the same. 

\begin{figure}[h]   
	\includegraphics[width=.99\textwidth]{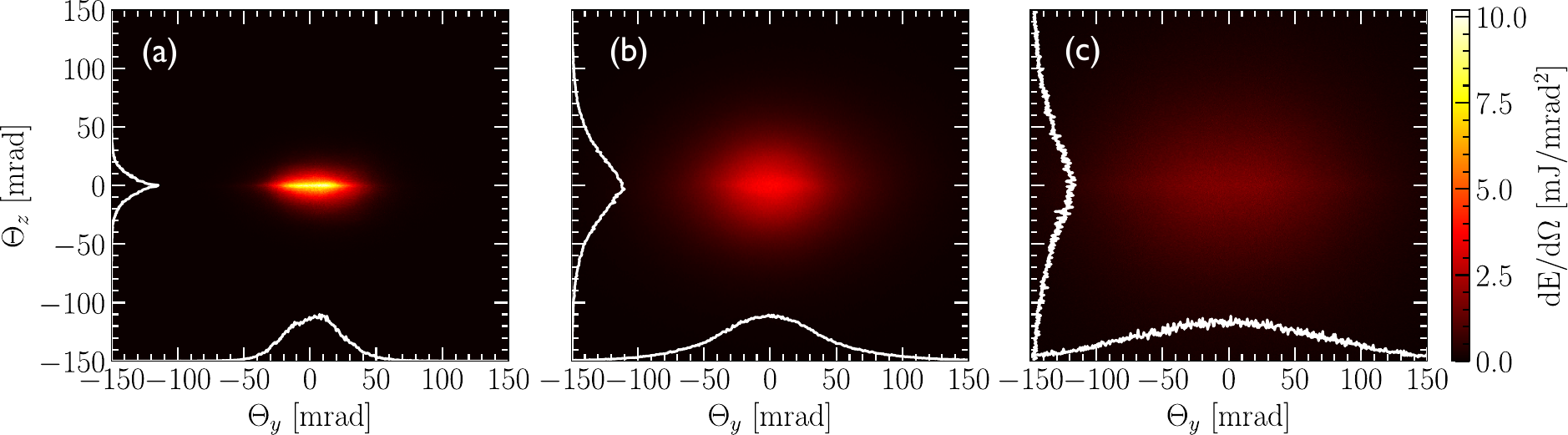}
	\caption{Divergence of emitted radiation detected on a far-field detector varies with the plasma density and the laser intensity and is in quantitative agreement with Eq. (\ref{eq:theta}).  (a) $n_e =0.01~ n_c$, $a_0 = 48$ (b) $0.05~ n_c$, $a_0 = 60$ and (c) $0.2~n_c$, $a_0 = 96$. All three boxes represent the interaction of a 5 PW laser. }
	\label{fig:divergence}
\end{figure}

The value of background plasma density might not have an impact on the value of critical frequency in optimal conditions, but it influences how collimated the radiation is. This is demonstrated in Fig. \ref{fig:divergence}, which shows the emission recorded on the far field detector. Three compared cases correspond to the radiation emitted during the interaction of 5 PW laser with targets with plasma densities of 0.01 $n_c$, 0.05 $n_c$ and 0.2 $n_c$ and laser amplitude $a_0$ of 48, 60 and 96 respectively. We observe that the divergence of emitted radiation improves with decreasing density despite slightly lowering laser intensity due to wider focusing at lower densities. The divergence is in excellent agreement with Eq. (\ref{eq:theta}), that predicts the divergence to be 34 mrad, 55 mrad and 74 mrad (for densities of 0.01 $n_c$, 0.05 $n_c$ and 0.2 $n_c$). Values fitted from simulations were 34 mrad, 68 mrad and 74 mrad.  One could expect that lower density results in significantly lower photon yield because of the dependence of emitted photons on the strength of the field. However, lower densities enable longer laser propagation due to slower laser depletion, which compensates for lower instantaneous photon emission rate. We observe the radiation with comparable conversion efficiencies from laser to photons in all three cases, namely 7 \% for the density 0.2 $n_c$, 5 \% for the density 0.05 $n_c$ and 1.5 \% for the density 0.01 $n_c$. The interaction with the background density of $0.01 ~n_c$ resulted in the brightest emission despite the lowest total radiated energy because of the divergence improving with the lower background plasma densities, see Fig. \ref{fig:divergence}. 

The brilliance can be also estimated from the simulation results. For example, for the case of a 5 PW laser propagating through the plasma background density $n_e = 0.05 ~n_c$, the critical energy was estimated as 58 MeV, with $10^9$ photons per 0.1\% of the bandwidth emitted at this energy. The radiation divergence (FWHM) was 32 mrad in the direction out of laser polarization plane and 68 mrad in the polarization plane, while the source was a cylinder with a radius of 3 $\rm{\mu}$m. The duration of the emitted photon radiation was estimated from the diagnostics as 10 fs. This results in a brilliance of $3\times 10^{22} $ photons / $\rm{mm^2mrad^2s}$ 0.1\% BW at the radiation critical energy 58 MeV. 

\section{Conclusion}

We have shown the potential of the DLA electrons as a collimated source of high-brilliance radiation. At first, we have verified the optimal focusing strategy that results in the highest electron cut-off energies achievable in the DLA regime. We observe that the multi-petawatt lasers are capable of accelerating electrons to energies exceeding 1 GeV limit, with 10 PW laser acceleration electrons up to 7.5 GeV. Energies achieved in simulations are in very good agreement with the theoretical prediction. Even though the simulations were performed using preformed guiding parabolic channels, it is possible to achieve nearly optimal acceleration also in non-preformed plasmas, where laser guiding relies on relativistic self-focusing\cite{babjak2024}. Electrons being accelerated in focusing ion channel field perform betatron oscillations, which results in synchrotron-like emission in the electron propagation direction. We show that critical energies of the emitted radiation range from several MeV for 0.1 PW lasers up to over 100 MeV for 10 PW lasers. Such high photon energies can be reached because of the maximization of the electron energy. Furthermore, the high energy of radiating electrons helps to improve the divergence of emission, resulting in the collimation of the radiated energy in a few tens of milirad. The total accelerated charge exceeding 100 nC results in a high yield of photons that for the most of the runs reached value around $10^8-10^{10}$ photons / 0.1\% bandwidth around the critical frequency. The combination of such high photon numbers, good divergence, small source size and short photon beam duration in tens of fs results in the brilliance in order of $10^{22} $ photons / $\rm{mm^2mrad^2s}$ 0.1\% BW at tens of MeV. Radiation source with such properties has a potential in many applications such as neutron generation, high-density states of plasma diagnostics or nuclear physics. 

\begin{acknowledgments}
 This work was supported by FCT Grants PTDC/FIS-PLA/3800/2021 DOI: https://doi.org/10.54499/PTDC/FIS-PLA/3800/2021 and FCT UI/BD/151560/2021 DOI: https://doi.org/10.54499/UI/BD/151560/2021. We acknowledge use of the LUMI (Finland) and Karolina (Czechia) supercomputers through EuroHPC awards. This work was supported by the Ministry of Education, Youth and Sports of the Czech Republic through the e-INFRA CZ (ID:90254).
\end{acknowledgments}

%\appendix

%\section{app 1}

%test

%\nocite{*}
\bibliography{bibliography}% Produces the bibliography via BibTeX.

\end{document}